\documentclass{article}
\usepackage{spconf,amsmath,graphicx,amssymb}
\usepackage{cite}
\usepackage{hyperref}
\hypersetup{
    colorlinks=true,
    linkcolor=blue,
    filecolor=magenta,      
    urlcolor=cyan,
    citecolor=blue,
    }
\usepackage{upgreek}
\title{Sample Abundance for Signal Processing: A Brief Introduction}
\name{Arian Eamaz, Farhang Yeganegi, Mojtaba Soltanalian}
\address{Department of Electrical and Computer Engineering\\ University of Illinois Chicago}
\usepackage{amsmath}
\usepackage{algpseudocode}
\usepackage{amsthm}
\usepackage{dsfont}
\usepackage{algorithm}
\usepackage{thmtools}
\usepackage{amssymb}
\usepackage[dvipsnames]{xcolor}

\def\bGamma{\boldsymbol{\Gamma}}

\def\bOmega{\boldsymbol{\Omega}}

\def\mba{\mathbf{a}}
\def\mbb{\mathbf{b}}
\def\mbc{\mathbf{c}}

\def\mbr{\mathbf{r}}

\def\mbx{\mathbf{x}}
\def\mby{\mathbf{y}}

\def\mbA{\mathbf{A}}
\def\mbB{\mathbf{B}}
\def\mbC{\mathbf{C}}

\def\mbP{\mathbf{P}}

\def\mbR{\mathbf{R}}

\def\mbX{\mathbf{X}}

\algnewcommand\algorithmicinput{\textbf{Input:}}
\algnewcommand\Input{\item[\algorithmicinput]}
\algnewcommand\algorithmicoutput{\textbf{Output:}}
\algnewcommand\Output{\item[\algorithmicoutput]}
\algnewcommand\algorithmicinit{\textbf{Initialize:}}
\algnewcommand\Init{\item[\algorithmicinit]}

\begin{document}
\ninept
\maketitle

\begin{abstract}
This paper reports, by way of introduction, on the advances made by our group and the broader signal processing community on the concept of \emph{sample abundance}; a phenomenon that naturally arises in one-bit and few-bit signal processing frameworks.
 By leveraging large volumes of low-precision measurements, we show how traditionally costly constraints—such as matrix semi-definiteness and rank conditions—become redundant, yielding simple overdetermined linear feasibility problems. We illustrate key algorithms, theoretical guarantees via the Finite Volume Property, and the \emph{sample abundance singularity} phenomenon, where computational complexity sharply drops.
\end{abstract}


\section{Introduction}
Sample abundance refers to the ability of modern signal processing systems, particularly those using one-bit analog-to-digital converters (ADCs), to collect a massive number of low-precision samples in a short time. Such abundance stems from extreme quantization of continuous signals into sign outputs ($\pm1$) with time-varying thresholds, enabling superior sampling rates, reduced power consumption, and low cost. 

The \emph{first key observation} is that sample abundance significantly helps in better observing and understanding the signals of interest. An old tale from Rumi would be helpful in this context \cite{javaherbin2015elephant}: \emph{
A group of villagers hears that a strange and enormous creature from India has arrived. Eager to understand it, they sneak into the dark barn where it's kept.
One touches the trunk and says, “It’s like a snake!”
Another, feeling a leg, declares, “No, it’s like a tree trunk!”
A third, touching an ear, insists, “It’s like a fan!”}
Who is right? The truth is, none of them sees the whole elephant. But what if they had more points of contact? Even if each touch lacked fine detail, the abundance of samples would provide a much better picture of the creature.  Analogously, one-bit sampling trades per-measurement precision for sheer volume, unlocking new signal insights.

The \emph{second key observation} is that under quantization, increasing data volume can reduce complexity in a counterintuitive manner. This is associated with what we refer to as the \emph{sample abundance singularity.} As a result of this phenomenon, many signal recovery tasks can be recast from constrained optimization to a simple linear feasibility test.

To properly introduce these ideas,  we will:
\begin{itemize}
  \item Define both the \emph{sample abundance} and the associated computational \emph{singularity}, where beyond a threshold of measurements, difficult constraints (e.g., positive semidefiniteness, signal rank) become unnecessary, leading to sharp drop in computational cost.
  \item Review enabling hardware (one-bit ADCs) and the role of time-varying thresholds, alongside a brief survey of foundational and recent developments in the domain.
  \item Present algorithms, particularly Randomized Kaczmarz algorithm (RKA) and sampling Kaczmarz-Motzkin method (SKM) variants, that solve the resulting overdetermined systems efficiently.
  \item Summarize the associated theoretical guarantees via the Finite Volume Property (FVP).

\end{itemize}

We begin by introducing the linear feasibility problems that naturally emerge in dithered one-bit quantization, across a range of signal processing tasks such as sparse parameter estimation \cite{thrampoulidis2020generalized}, phase retrieval \cite{eamaz2022phase}, compressed sensing \cite{xu2020quantized, dirksen2021non, Eamaz2023QuadraticCS}, low-rank matrix sensing \cite{eamaz2024harnessing,Yeganegi2024LowRankOneBit}, and sampling technology \cite{Eamaz2024UNO, Eamaz2024HDR,Eamaz2023UnlimitedSampling,Eamaz2025StreamliningUNO}. These problems lead to what we refer to as the one-bit polyhedron formulation consisting of a set of linear inequalities representing the feasible region for recovery.

To solve these linear feasibility problems, various randomized iterative algorithms have been proposed, many of which are variants of the RKA. Randomized methods have become a fundamental tool in high-dimensional data processing, providing efficient solutions for large-scale linear systems and their extensions to non-linear optimization problems \cite{martinsson2020randomized}. These include not only RKA \cite{strohmer2009randomized}, but also randomized coordinate descent methods \cite{leventhal2010randomized}. A key feature of these algorithms is that they use sketched or randomly selected subsets of the data at each iteration, rather than processing the entire dataset. This sketching approach allows for substantial reductions in computational complexity.

Beyond algorithm design, the theoretical analysis of sample complexity—i.e., the number of measurements required to achieve a desired recovery error—has been a central topic in the literature on dithered one-bit sampling. Since dithered one-bit systems introduce additional randomness via time-varying or randomly generated thresholds, they enable the development of statistical recovery guarantees, even under extremely coarse quantization.
A significant open problem in this field is the efficient generation of dithers. The goal is to design dithering schemes that do not impose additional computational or storage burdens during reconstruction. Further research is needed to develop more efficient and practical dithering methods, both in terms of implementation and theoretical understanding.

\emph{Notation:} We use bold lowercase letters for vectors and bold uppercase letters for matrices. The notation $\mathbb{R}$ represents the set of real numbers. We use $(\cdot)^{\top}$ to denote the vector/matrix transpose. The operator $\operatorname{Tr}(.)$ denotes the trace of the matrix argument. The standard inner product between two matrices is defined as $\left\langle \mbB_{1},\mbB_{2}\right\rangle=\operatorname{Tr}(\mbB_{1}^{\mathrm{H}}\mbB_{2})$. The nuclear norm of a matrix $\mbB\in \mathbb{R}^{N_{1}\times N_{2}}$ is denoted $\left\|\mbB\right\|_{\star}=\sum^{M}_{i=1}\sigma_{i}$ where $M$ and $\left\{\sigma_{i}\right\}$ are the rank and singular values of $\mbB$, respectively. The Frobenius norm of a matrix $\mbB$ is defined as $\|\mbB\|_{\mathrm{F}}=\sqrt{\sum^{N_{1}}_{r=1}\sum^{N_{2}}_{s=1}\left|b_{rs}\right|^{2}}$ where $\{b_{rs}\}$ are elements of $\mbB$. The $\ell_{k}$-norm of a vector $\mathbf{b}$ is defined as $\|\mbb\|^{k}_{k}=\sum_{i}|b|^{k}_{i}$. The Hadamard (element-wise) product of two matrices $\mbB_{1}$ and $\mbB_{2}$ is denoted as $\mbB_{1}\odot \mbB_{2}$. The vectorized form of a matrix $\mbB$ is written as $\operatorname{vec}(\mbB)$. If there exists a $c>0$ such that $a\geq c b$ (resp. $a\leq c b$) for two quantities $a$ and $b$, we have $a \gtrsim b$ (resp. $a \lesssim b$). The set $[n]$ is defined as $[n]=\left\{1,\cdots,n\right\}$. A covering number is the number of $r$-balls of a given size needed to completely cover a given set $\mathcal{K}$, i.e., $\mathcal{N}\left(\mathcal{K},\|\cdot\|_2, r\right)$. The Kolmogorov $r$-entropy of a set $\mathcal{K}$ is denoted by $\mathcal{H}\left(\mathcal{K},r\right)$ defined as the logarithm of the size of the smallest $r$-net of $\mathcal{K}$.

\vspace{-10pt}
\section{One-Bit ADCs and Sample Abundance}
Early work has already demonstrated that even extreme one-bit and few-bit quantization, when coupled with time-varying thresholds, can support high-fidelity parameter estimation \cite{thrampoulidis2020generalized}. The re-emergence of one-bit and few-bit ADCs is, however, mostly indebted to the growing interest in massive MIMO communications\cite{mezghani2018blind}, where a large number of RF chains would require the same amount of ADCs in the system architecture. Due to their simple setup, effectively comparators, one-bit ADCs (and few-bit variants) have a significant advantage when it comes to sampling rate, power consumption, and cost, which is why they have been considered as a solution in the massive MIMO context.

Many applications of one-bit ADCs require the estimation of the signal statistics to pave the way for parameter estimation. On the statistical side,
\cite{eamaz2023covariance} extended the classical arcsine law\cite{van2005spectrum} to one-bit stationary signals with drifting thresholds. This work was further generalized to non-stationary processes in \cite{Eamaz2022CovarianceNonStationaryTSP}. These results enable accurate covariance and spectrum estimation directly from sign data. Based on these statistical foundations, \cite{Ameri2018OneBitRadarSAM} introduces weighted least-squares estimation for radar target returns from binary samples with sweeping thresholds, followed by complete statistical characterizations and Cramér–Rao bounds for one-bit radar processing \cite{Ameri2019OneBitRadarTSP}. These pioneering studies confirm that low-cost comparators could achieve near-ideal performance at high sampling rates.

Building on this, \cite{Khobahi2018AdaptiveThresholding} proposed adaptive threshold updates within a constrained quadratic program to robustly recover signals under colored noise. Subsequent deep unfolding approaches, such as DeepRec \cite{Khobahi2019DeepRec} and LoRD-Net for few-bit MIMO detection \cite{Khobahi2021ModelInspired}, demonstrated that model-inspired neural architectures could dramatically accelerate low-resolution signal recovery.

The recent invention of \emph{unlimited one-bit sampling} (UNO) bridges modulo sampling with one-bit quantization. The efforts in  \cite{Eamaz2023UnlimitedSampling,Eamaz2024UNO,Eamaz2025StreamliningUNO} introduce hybrid ADC architectures that combine modulo folding and uniform random thresholds, achieving bandlimited-signal recovery with minimal hardware. Extensions include finite-rate-of-innovation streams \cite{Eamaz2023UnlimitedFRI} and high dynamic range (HDR) imaging via threshold sweeps \cite{Eamaz2024HDR}.

In applications where low-rank matrix recovery is useful, one-bit matrix completion with time-varying thresholds \cite{Eamaz2023MatrixCompletionThreshold,Eamaz2024MatrixCompletionDither} and structured Hankel matrix recovery \cite{Eamaz2024AutoRadarHankel} have been studied, showing that coarse binary data can reliably reconstruct high-dimensional matrices under moderate oversampling.

Collectively, these efforts underscore that one-bit ADCs, empowered by time-varying thresholds, unlock \emph{sample abundance}---the ability to gather a relatively large number of sign measurements per second. This abundance compensates for coarse quantization, laying the groundwork for novel reconstruction frameworks.

\begin{figure*}[t]
  \centering
  \includegraphics[width=0.8\linewidth]{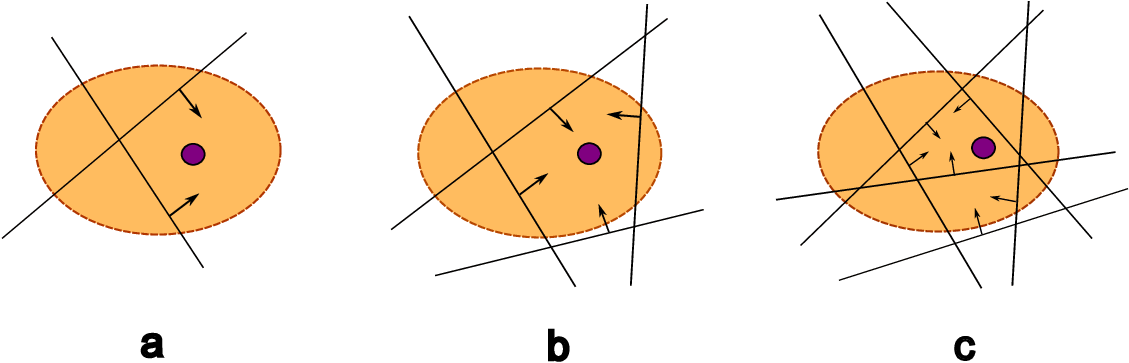}
   \caption{Shrinkage of the one-bit polyhedron, ultimately placed within the constraint space shown with the orange region, when the number of constraints (samples) grows large. The arrows point to the half-space associated with each inequality constraint. The evolution of the feasible regime is depicted with increasing samples in three cases: (a) small sample-size regime, constraints not forming a finite-value polyhedron; (b) medium sample-size regime, constraints forming a finite-volume polyhedron, parts of which are outside the constraint space; (c) large sample-size regime, constraints forming a finite-volume polyhedron inside the orange region, making its constraint redundant.}
  \label{fig:abundance}
  \vspace{-10pt}
\end{figure*}

\section{From Complex Optimization to Linear Feasibility Tests}
Let $y_{k}=y(k\mathrm{T})$ denote the uniform samples of signal $y(t)$ with the sampling rate $1/\mathrm{T}$. In practice, the discrete-time samples occupy pre-determined quantized values. We denote the quantization operation on $y_{k}$ by the function $Q(\cdot)$. This yields the scalar quantized signal as $r_{k} = Q(y_{k})$.
In one-bit quantization, compared to zero or constant thresholds, time-varying sampling thresholds yield a better recovery performance \cite{eamaz2023covariance}. These thresholds may be chosen from any distribution.  
In the case of one-bit quantization with such time-varying sampling thresholds, we have
$r_{k} = \operatorname{sgn}\left(y_{k}-\tau_{k}\right)$.
The information gathered through the one-bit sampling with time-varying thresholds presented here may be formulated in terms of an overdetermined linear system of inequalities. 
We have $r_{k}=+1$ when $y_{k}>\tau_{k}$ and $r_{k}=-1$ when $y_{k}<\tau_{k}$. Therefore, one can formulate the geometric location of the signal as
$r_{k}\left(y_{k}-\tau_{k}\right) \geq 0$.
Collecting all the elements in the vectors $\mathbf{y}=[y_{k}] \in \mathbb{R}^{n}$ and $\mathbf{r}=[r_{k}] \in \left\{-1,1\right\}^{n}$, we have
$\mbr\odot\left(\mathbf{y}-\boldsymbol{\tau}\right) \succeq \mathbf{0}$,
or equivalently
\begin{equation}
\label{eq:6}
\begin{aligned}
\bOmega_{\mby}  \mathbf{y} &\succeq \mbr \odot \boldsymbol{\tau},
\end{aligned}
\end{equation}
where $\bOmega_{\mby}  \triangleq \operatorname{diag}\left\{\mbr\right\}$. Denote the time-varying sampling threshold in $\ell$-th signal sequence by $\boldsymbol{\tau}^{(\ell)}$, where  $\ell\in [m]$.
It follows from \eqref{eq:6} that
\begin{equation}
\label{eq:7}
\begin{aligned}
\bOmega^{(\ell)}_{\mby} \mathbf{y} &\succeq \mathbf{r}^{(\ell)} \odot \boldsymbol{\tau}^{(\ell)}, \quad \ell \in [m],
\end{aligned}
\end{equation}
where $\bOmega^{(\ell)}_{\mby} =\operatorname{diag}\left(\mbr^{(\ell)}\right)$. 
Denote the concatenation of all $m$ sign matrices as
\begin{equation}
\label{eq:9}
\Tilde{\bOmega}_{\mby} =\left[\begin{array}{c|c|c}
\bOmega^{(1)}_{\mby}  &\cdots &\bOmega^{(m)}_{\mby} 
\end{array}\right]^{\top}, \quad \Tilde{\bOmega}_{\mby} \in \left\{-1,0,1\right\}^{m n\times n}.
\end{equation}
Rewrite the $m$ linear 
inequalities in \eqref{eq:7} as
\begin{equation}
\label{eq:8}
\Tilde{\bOmega}_{\mby}  \mathbf{y} \succeq \operatorname{vec}\left(\mbR_{\mby} \right)\odot \operatorname{vec}\left(\bGamma\right),
\end{equation}
where $\mathbf{R}$ and $\bGamma$ are matrices, whose columns are the sequences $\left\{\mathbf{r}^{(\ell)}\right\}_{\ell=1}^{m}$ and $\left\{\boldsymbol{\tau}^{(\ell)}\right\}_{\ell=1}^{m}$, respectively.

Assuming a large number of samples --- a common situation in one-bit sampling scenarios --- hereafter we treat (\ref{eq:8}) as an overdetermined linear system of inequalities associated with the one-bit sensing scheme.
The inequality (\ref{eq:8}) can be recast as a polyhedron,
\begin{equation}
\label{eq:8n}
\begin{aligned}
\mathcal{P}_{\mby}=\left\{\mby^{\prime} \in \mathbb{R}^n \mid \tilde{\boldsymbol{\Omega}}_{\mby} \mby^{\prime} \succeq \operatorname{vec}\left(\mbR_{\mby}\right) \odot \operatorname{vec}(\boldsymbol{\Gamma})\right\} \subset \mathbb{R}^n,
\end{aligned}
\end{equation}
which we refer to as the \emph{one-bit polyhedron}. Generally, it can be assumed that the signal $\mbx\in\mathbb{R}^{d}$ is observed linearly through the sampling matrix $\mbA\in\mathbb{R}^{n\times d}$, creating the measurements as $\mby=\mbA\mbx$. Based on \eqref{eq:8}, the one-bit polyhedron for this type of problem is given by
\begin{equation}
\label{eq:80n}
\begin{aligned}
\mathcal{P}_{\mathbf{x}} = \left\{\mathbf{x}^{\prime}\in\mathbb{R}^{d} \mid \mbP_{\mby}  \mathbf{x}^{\prime} \succeq \operatorname{vec}\left(\mbR_{\mby} \right)\odot \operatorname{vec}\left(\bGamma\right)\right\}\subset \mathbb{R}^d,
\end{aligned}
\end{equation}
where $\mbP_{\mby} =\Tilde{\bOmega}_{\mby} \mbA$, or equivalently
\begin{equation}
\label{eq:90}
\mbP_{\mby} =\left[\begin{array}{c|c|c}
\mbA^{\top}\bOmega^{(1)}_{\mby}&\cdots &\mbA^{\top}\bOmega^{(m)}_{\mby} 
\end{array}\right]^{\top}, \quad \mbP_{\mby} \in\mathbb{R}^{m n\times d}.
\end{equation}
In the following, we present the linear feasibility formulations emerging in classical problems such as phase retrieval, compressed sensing, and low-rank matrix sensing.
\\
$\bullet$ \textbf{\emph{Phase Retrieval:}}
Classical recovery methods for magnitude-only or low-bit quantized data rely on \emph{lifting} to enforce non-convex constraints. For instance, PhaseLift casts phase retrieval as a semi-definite program (SDP) by embedding the vector signal into a rank-one positive semi-definite matrix \cite{Eamaz2022OneBitPhasePR}. While globally optimal, SDP solvers exhibit prohibitive memory and runtime costs, scaling superlinearly in signal dimension.

By contrast, with abundant one-bit measurements, we formulate recovery as a system of linear inequalities: 
\begin{equation}
\label{eq_pr}
\begin{aligned}
\mathcal{P}_{\mbX}^{(P)} = \left\{\mbX^{\prime}\in\mathbb{R}^{n_1\times n_2} \mid r_j^{(\ell)}\mba_j^{\top}\mbX^{\prime}\mba_j\succeq r_j^{(\ell)}\tau_j^{(\ell)}\right\}\subset\mathbb{R}^{n_1\times n_2},
\end{aligned}
\end{equation} 
where each measurement $r_j^{(\ell)} = \mathrm{sign}\left(\mba_j^{\top}\mbX\mba_j-\tau_j^{(\ell)}\right)$ for $j\in[m],\ell\in[L]$. The intersection of $m$ half-spaces forms a polyhedron enclosing the true signal. As $m$ increases, this polyhedron shrinks toward $\mbX$, rendering PSD and rank constraints redundant beyond a sample threshold \cite{Eamaz2024SampleAbundanceIT,Eamaz2023QuadraticCS}.

This paradigm shift yields two key advantages:
\begin{itemize}
  \item \textbf{Scalability}: Projection-based solvers for linear feasibility scale linearly in $m$ and $n$, contrasting sharply with cubic SDP complexity.
  \item \textbf{Simplicity}: The finite volume created by one-bit samples around the desired solution reduces memory footprint and removes the need for specialized SDP solvers, enabling real-time implementations.
\end{itemize}
$\bullet$ \textbf{\emph{Low-rank Matrix Sensing:}} The low-rank matrix sensing problem can be formulated as:
\begin{equation}
\label{eq:1nnnnn}
\begin{aligned}
\text{find}\quad \mbX \in \Omega_{c} \quad
\text{subject to} \quad \mathcal{A}\left(\mbX\right)=\mathbf{y},~ 
\operatorname{rank}\left(\mbX\right)\leq r,
\end{aligned}
\end{equation}
where $\mbX\in \mathbb{R}^{n_{1}\times n_{2}}$ is the matrix of unknowns, $\mathbf{y}\in \mathbb{R}^{m}$ is the measurement vector, and $\mathcal{A}$ is a linear transformation such that $\mathcal{A}:\mathbb{R}^{n_1\times n_2}\mapsto\mathbb{R}^m$.
In general, $\Omega_{c}$ can be chosen such as the set of semi-definite matrices, symmetric matrices, upper or lower triangle matrices, Hessenberg matrices and a specific constraint on the matrix elements $\left\|\mbX\right\|_{\infty}\leq \alpha$ or on its eigenvalues, i.e., $\lambda_{i}\leq \epsilon$ where $\left\{\lambda_{i}\right\}$ are eigenvalues of $\mbX$. The nuclear norm minimization alternative of the problem is given by:
\begin{equation}
\label{eq:1nnnnnnn}
\begin{aligned}
\underset{\mbX \in \Omega_{c}}{\textrm{minimize}}\quad \left\|\mbX\right\|_{\star}\quad
\text{subject to} \quad \mathcal{A}\left(\mbX\right)=\mathbf{y}.
\end{aligned}
\end{equation}
In the context of low-rank matrix sensing, the linear operator $\mathcal{A}\left(\mbX\right)$ is typically defined as:
\begin{equation}
\label{Stefanie_2}
\mathcal{A}\left(\mbX\right)=\frac{1}{\sqrt{m}}\left[\operatorname{Tr}\left(\mbA^{\top}_1\mbX\right)\cdots\operatorname{Tr}\left(\mbA^{\top}_m\mbX\right)\right]^{\top},
\end{equation}
where $\mbA_j\in\mathbb{R}^{n_1\times n_2}$ is the $j$-th sensing matrix. The one-bit polyhedron associated with the low-rank matrix sensing setup is given by:
\begin{equation}
\label{kumar}
\mathcal{P}_{\mbX}^{(M)}=\left\{\mbX^{\prime}\in\mathbb{R}^{n_1\times n_2} \mid r^{(\ell)}_{j}y_j\geq r^{(\ell)}_{j}\tau^{(\ell)}_{j}\right\}\subset \mathbb{R}^{n_1\times n_2},
\end{equation}
where $y_j=\operatorname{Tr}\left(\mbA^{\top}_j\mbX\right)/\sqrt{m}$, and each binary measurement is given by $r_j^{(\ell)} = \mathrm{sign}\left(y_j-\tau_j^{(\ell)}\right)$ for $j\in[m],\ell\in[L]$.\\
$\bullet$ \textbf{\emph{Compressed Sensing:}} The compressed sensing problem is formulated as:
\begin{equation}
\label{1}
\begin{aligned}
\underset{\mbx \in \mathbb{R}^n}{\textrm{minimize}}\quad \left\|\mbx\right\|_{0}\quad
\text{subject to} \quad \mby=\mbA\mbx,
\end{aligned}
\end{equation}
where $\mbx\in\mathbb{R}^n$ is a sparse unknown signal, $\mbA\in\mathbb{R}^{m\times n}$ is the sensing matrix, and $\mby\in\mathbb{R}^m$ is the measurement vector. Since the optimization problem in \eqref{1} is NP-hard due to the non-convex $\ell_0$-norm, it is commonly relaxed using the convex $\ell_1$-norm minimization:
\begin{equation}
\label{2}
\begin{aligned}
\underset{\mbx \in \mathbb{R}^n}{\textrm{minimize}}\quad \left\|\mbx\right\|_{1}\quad
\text{subject to} \quad \mby=\mbA\mbx.
\end{aligned}
\end{equation}
In the one-bit compressed sensing setting, recovery is achieved by constructing a highly constrained one-bit polyhedron based on a collection of binary measurements:
\begin{equation}
\label{Stefanie_500}
\mathcal{P}_{\mbx}^{(C)}=\left\{\mbx^{\prime}\in \mathbb{R}^{n}\mid r^{(\ell)}_{j}\langle\mba_j,\mbx^{\prime}\rangle\geq r^{(\ell)}_{j}\tau^{(\ell)}_{j}\right\}\subset \mathbb{R}^{n},
\end{equation}
where each binary measurement is given by $r_j^{(\ell)} = \mathrm{sign}\left(\langle\mba_j,\mbx\rangle-\tau_j^{(\ell)}\right)$ for $j\in[m],\ell\in[L]$.
 
The space formed by the intersection of linear constraints can fully shrink to the desired signal $\mbx$ inside the feasible region, which is shown by the orange space---see Fig.~\ref{fig:abundance} for an illustrative example of this phenomenon. As can be seen in this figure, the blue lines displaying the linear feasibility form a finite-volume space around the original signal displayed by the purple circle inside the elliptical region by growing the number of threshold sequences or one-bit samples. In (a), constraints are not enough to create a finite-volume space, whereas in (b) such constraints can create the desired finite-volume polyhedron space which, however, is not fully inside the orange space. Lastly, in (c), the created finite-volume space shrinks to be fully inside the feasible region. 

\section{Algorithms Leveraging Sample Abundance}
Iterative projection methods form the required computational backbone:
\begin{itemize}
  \item \textbf{RKA}: The RKA is a \emph{sub-conjugate gradient method} to solve a linear feasibility problem, i.e, $\mbC\mathbf{x}\succeq\mathbf{b}$ where $\mbC$ is a ${m\times n}$ matrix with $m>n$ \cite{leventhal2010randomized,strohmer2009randomized}. Conjugate-gradient methods immediately turn the mentioned inequality to an equality in the following form $\left(\mbb-\mbC\mathbf{x}\right)^{+}=0$, and then, approach the solution by the same process as used for systems of equations. The projection coefficient $\beta_{i}$ of the RKA is \cite{leventhal2010randomized}:
  \begin{equation}
  \label{eq:22}
  \beta_{i}= \begin{cases}
  \left(b_{j}-\langle\mathbf{c}_{j},\mathbf{x}_{i}\rangle\right)^{+} & \left(j \in \mathcal{I}_{\geq}\right), \\ b_{j}-\langle\mathbf{c}_{j},\mathbf{x}_{i}\rangle & \left(j \in \mathcal{I}_{=}\right),
  \end{cases}
  \end{equation}
  where the disjoint index sets $\mathcal{I}_{\geq}$ and $\mathcal{I}_{=}$ partition $[m]$ and $\{\mathbf{c}_{j}\}$ are the rows of $\mathbf{C}$. Also, the unknown column vector $\mathbf{x}$ is iteratively updated as
  \begin{equation}
  \label{eq:23}
  \mathbf{x}_{i+1}=\mathbf{x}_{i}+\frac{\beta_{i}}{\left\|\mbc_{j}\right\|^{2}_{2}} \mbc^{\star}_{j},
  \end{equation}
  where, at each iteration $i$, the index $j$ is drawn from the set $[m]$ independently at random following the distribution
  $\operatorname{Pr}\{j=k\}=\frac{\left\|\mbc_{k}\right\|^{2}_{2}}{\|\mbC\|_{\mathrm{F}}^{2}}$. Assuming that the linear system is consistent with nonempty feasible set $\mathcal{P}_{\mbx}$ created by the intersection of hyperplanes around the desired point $\mbx$, RKA converges linearly in expectation to the solution $\widehat{\mbx}\in\mathcal{P}_{\mbx}$\cite{strohmer2009randomized,leventhal2010randomized}:
  \begin{equation}
  \label{eq:15}
  \mathbb{E}\left\{\hbar\left(\mathbf{x}_{i},\widehat{\mbx}\right)\right\} \leq \left(1-q_{_{\text{RKA}}}\right)^{i}~ \hbar\left(\mathbf{x}_{0},\widehat{\mbx}\right),
  \end{equation}
  where $\hbar\left(\mathbf{x}_{i},\widehat{\mbx}\right)=\left\|\mathbf{x}_{i}-\widehat{\mbx}\right\|_{2}^{2}$ is the euclidean distance between two points in the space, $i$ is the number of required iterations for RKA, and $q_{_{\text{RKA}}} \in \left(0,1\right)$ is given by $q_{_{\text{RKA}}}=\frac{1}{\kappa^{2}\left(\mbC\right)}$, with $\kappa\left(\mbC\right)=\|\mbC\|_{\mathrm{F}}\|\mbC^{\dagger}\|_{2}$ denoting the scaled condition number of a matrix $\mbC$. 
  \item \textbf{SKM}: The SKM combines the ideas of both the RKA and the Motzkin method. The generalized convergence analysis of the SKM with sketch matrix which has been formulated based on the convergence analysis of RKA, and sampling Motzkin method for solving linear feasibility problem has been comprehensively explored in \cite{de2017sampling}. The central contribution of SKM lies in its innovative way of projection plane selection. The hyperplane selection is done as follows: At iteration $i$, the SKM algorithm selects a collection of $\gamma$ (denoted by the set $\mathcal{T}_{i}$) rows, uniformly at random out of $m$ rows of the constraint matrix $\mbC$. Then, out of these $\gamma$ rows, the row $j_{\star}$ with the maximum positive residual is selected; i.e.
  \begin{equation}
  j_{\star}=\operatorname{argmax}~\left\{ \left(b_{j}-\langle\mbc_{j},\mathbf{x}_{i}\rangle\right)^{+}\right\},~j\in\mathcal{T}_{i}.
  \end{equation}
  Finally, the solution is updated as $\mathbf{x}_{i+1}=\mathbf{x}_{i}+\lambda_{i}\frac{\beta_{i}}{\left\|\mbc_{j_\star}\right\|^{2}_{2}} \mbc^{\star}_{j_{\star}}$ \cite{de2017sampling}, where $\lambda_{i}$ is a relaxation parameter which for consistent systems must satisfy $0\leq \lim_{i\to\infty} \inf \lambda_{i}\leq \lim_{i\to\infty} \sup \lambda_{i}<2$, to ensure convergence \cite{strohmer2009randomized}.
\end{itemize}
These algorithms bring about the \emph{sample abundance singularity}: once enough measurements are collected, the number of iterations to achieve target accuracy drops precipitously, offsetting the overhead of additional samples.

\section{Finite Volume Property}
The Finite Volume Property (FVP) formalizes how the volume of the polyhedron defined by one-bit inequalities decays with the number of measurements. For isotropic sensing matrices, FVP guarantees that confining the feasible set within an $\epsilon$-ball requires
\[ m = O(\epsilon^{-3}), \]
yielding a reconstruction error scaling of $O(m^{-1/3})$ \cite{Eamaz2024SampleAbundanceIT,Yeganegi2024LowRankOneBit}.

Beyond unstructured Gaussian models, FVP extends to structured transforms (e.g., DCT, Fourier) under mild isotropy conditions, enabling guarantees, e.g., for co-array DOA estimation \cite{Sedighi2021OneBitCoArray} and non-stationary covariance recovery \cite{Eamaz2022CovarianceNonStationaryTSP}.

In the FVP theorems, we demonstrated \cite{eamaz2024harnessing} that incorporating structured sets (i.e., exploiting constraints) reduces the number of measurements required to achieve acceptable recovery performance. 
However, important questions remain: How does this sample abundance theoretically impact the recovery error? And at what rate does the error decay as the number of measurements increases? 

Without assuming any structure for the set, Kolmogorov $\rho$-entropy is bounded using Sudakov’s minoration,
\begin{equation}
m\gtrsim \frac{\gamma^2\left(\mathcal{K}\right)}{\rho^2 \epsilon^2},
\end{equation}
which has been demonstrated to be non-tight for structured sets \cite{oymak2015near}. For structured sets such as the low-rank matrix set, the upper bound on Kolmogorov $\rho$-entropy is constrained as follows \cite[Table~1]{jacques2017time}:
\begin{equation}
\mathcal{H}\left(\mathcal{K}_r,\rho\right) \lesssim  r \left(n_1+n_2\right)\log\left(1+\frac{1}{\rho}\right),
\end{equation}
and for sparse set
\begin{equation}
\mathcal{H}\left(\mathcal{K}_s,\rho\right) \lesssim  s\log\left(\frac{d}{s}\right)\log\left(1+\frac{1}{\rho}\right),
\end{equation}
which leads to
\[ m = O(\epsilon^{-2}). \]
These results provide mathematical credence to
the \emph{sample abundance singularity}: once $m$ surpasses the FVP threshold for a desired error, sample abundance obviates the need for taking complex PSD/rank constraints into account, precipitating a drop in computational cost.

\section{Conclusion}
We have surveyed how one-bit ADCs with time-varying thresholds enable \emph{sample abundance}, shifting recovery from complex optimization to scalable linear feasibility. Key projection algorithms (RKA, SKM) leverage abundance to achieve efficient reconstruction, underpinned by FVP theoretical guarantees. Future work will explore adaptive threshold trajectories, structured measurement extensions, and real-time hardware realizations.

\bibliographystyle{ieeetr}
\bibliography{master_onebit}

\end{document}